\documentclass[12pt]{article} 
\usepackage{epsfig}

\makeatletter 
\@addtoreset{equation}{section} 
\makeatother 
 
\newcommand{\beq}{\begin{equation}} 
\newcommand{\eeq}{\end{equation}} 
\newcommand{\be}{\begin{eqnarray}}
 \newcommand{\ee}{\end{eqnarray}}

\newcommand{\de}{\partial}

\newcommand{\ov } {\over } 

 \newcommand{\s }{\sigma }

\def\I{ {\rm I}}
\def\II{ {\rm II}}
\def\td{\tilde }
\def\a{\alpha }

\def\appendix#1{   \addtocounter{section}{1}   \setcounter{equation}{0}   
\renewcommand{\thesection}{\Alph{section}}   \section*{Appendix \thesection\protect\indent \parbox[t]{11.15cm}   {#1} }   \addcontentsline{toc}{section}{Appendix \thesection\ \ \ #1}   } 

\begin{document}

\begin{titlepage}

\begin{center}
\hfill hep-th/0301109\\
\hfill CERN-TH/2003-005\\
\hfill SISSA 5/2003/EP \\

\vskip  1 cm

\vskip 1 cm

{\Large \bf Semiclassical decay  of strings with maximum angular momentum}

\vskip  1 cm

{\large Roberto Iengo $^a$ and Jorge G. Russo$^{b,c}$}


\end{center}


\centerline{\it ${}^a$ International School for Advanced Studies (SISSA), Via Beirut 2-4,}
\centerline{\it I-34013 Trieste, Italy}
\centerline{\it  INFN, Sezione di Trieste }
\medskip

\centerline{\it {}$^b$ Theory Division, CERN, CH-1211 Geneva, Switzerland}

\medskip

\centerline {\it {}$^c$ Departamento de F\'\i sica, Universidad de
Buenos Aires and Conicet, }
\smallskip
\centerline {\it Ciudad Universitaria, Pab. I, 1428
Buenos Aires, Argentina}

\vskip 0.4 cm

\vskip 1.5 cm

\begin{abstract}

We study the classical breaking of 
a highly excited (closed or open) string state on the leading Regge trajectory,
represented by a rotating soliton solution, 
and we find the resulting solutions for the outgoing two pieces, 
describing two specific excited string states. 
This classical picture  reproduces very accurately 
the precise analytical relation of the masses $M_1$ and $M_2$ of the decay
products found in a previous quantum computation.  
The decay rate
is naturally described in terms of a semiclassical formula.
We also point out some interesting features of the evolution  after
the  splitting process.

\end{abstract}

\bigskip
\bigskip
\bigskip

\date{January 2003}

\end{titlepage}

\newpage


\noindent
{\bf Contents
\bigskip

\noindent 1. Introduction

\bigskip

\noindent 2. Splitting of a rotating string}

\medskip

2.1 Fourier analysis

\medskip

2.2 Closed formulas for the outgoing string solutions

\bigskip

{\bf

\noindent 3. Comparison with the superstring quantum calculation

\bigskip

\noindent 4. Decay rate and description of the motion -- 

\noindent ~~ Figures for the string splitting and evolution }

\bigskip
\bigskip

\section{Introduction}

The study of the decay properties of massive string states, and, in particular, 
the attempt to determine their
lifetime, was initiated long ago, with various kinds of  results   
\cite{Green,Mitchell,Dai,Okada,Sundborg,wilkinson,Turok,Amati2,IK,Manes,IR}
sometimes finding a long lifetime (see e.g. \cite{Green,wilkinson}). 
The main difficulty is represented by the high degeneracy of the decay products,
even for the splitting into two bodies, which can be viewed
as the first step of the decay process.

Here we present a simple classical computation, which is applicable
for large quantum numbers, describing the splitting into two pieces of 
a very massive closed string in the state of maximal, and thus very large, 
angular momentum (a study of classical 
splitting of a different circular pulsating string is in \cite{devega}). 
This computation will prescribe a definite relation between the 
masses of the decay products and will also give information about their angular momentum.
Moreover, the semiclassical argument indicates that the lifetime for the splitting
of such highly excited states will grow proportional to its mass.

We will also compare the results of this semiclassical picture  with the detailed
quantum computation that we recently performed on the decay of a
 massive state with maximal angular momentum in closed superstring theory  \cite{IR}.
The quantum computation
was based on evaluating the imaginary part of the one-loop self-energy of that state.
The self-energy is expressed as an integral of a certain combination of theta functions,
and we developed a very efficient algorithm to derive the contribution of 
the intermediate states of definite mass, which included the sum over their degeneracies.
  

The comparison will show a surprisingly good agreement 
between the semiclasssical and quantum calculation,
despite the fact they involve completely different computations.


\section{Splitting of a rotating string }

The rotating closed string solution is given by:
\beq
X=L \cos(2\s ) \cos(2 \tau) \ ,\ \ \ Y=L\cos(2 \s )\sin(2\tau ) \ ,\ \ \ \ X^0=2L\tau \ ,
\label{ccero}
\eeq
where $\s \in [0,\pi)$. This represents a spinning folded closed string with maximum angular momentum.
The coefficient $2L$ in $X^0$ is fixed  by the constraint
$\dot X \cdot \dot X + {X'\cdot X'}=0$ (the other constraint $\dot X\cdot X'=0$ is also satisfied).
The string has energy and angular momentum given by
\beq
E=M={L\over\a '}\ ,\ \ \ \ \ J={L^2\over 2\a'}\ ,
\eeq
so that one has the usual Regge relation $\a'M^2=2J$.

The closed string (\ref{ccero}) contains two segments, the ``upper'' 
segment described by $0<\s < {\pi\over 2}$ and the ``lower'' 
segment described by ${\pi\over 2} <\s < {\pi}$. The full length 
of the string is $4L$. We assume that at $\tau=0 $ the string 
splits into two pieces of lengths $4L_\I $ and $4L_{\II }$, 
$X^\mu (\s )\to \{ X^\mu_\I (\s ), X^\mu_{\II } (\s ) \} \ ,$ 
with $L_\I+L_\II=L$. The initial conditions for the strings I and 
II are given by the string (\ref{ccero}) at $\tau=0 $. 
 The splitting occurs at $\s ={a\pi \over 2}$
for the upper segment, with $a$ defined as
\beq
\cos a\pi \equiv -{L_\I-L_{\II}\over L}\ , \ \ \ \ 0<a<1\ .
\eeq
{}For the lower segment, the splitting is at $\s=\pi -{a\pi\over 2} $.
Thus the initial conditions are: 
\beq
X^\mu_\I (\s ,0)=X^\mu (\s, 0)\ ,\ \ \ 
\dot X^\mu_\I (\s ,0)=\dot X^\mu (\s, 0)\ ,\
\label{cdoz}
\eeq
for $ 0<\s< {a\pi\over 2}$ (upper segment) and 
for $\pi >\s> \pi -{a\pi\over 2} $ (lower segment),
while for the string II we have
\beq
X^\mu_{\II }(\s ,0)=X^\mu (\s, 0)\ ,\ \ \ 
\dot X^\mu_{\II }(\s ,0)=\dot X^\mu (\s, 0)\ ,\ 
\label{cunos}
\eeq
for ${a\pi\over 2} <\s< {\pi\over 2}$ (upper segment) and   
for ${\pi\over 2} <\s< \pi -{a\pi\over 2} $ (lower segment).
These boundary conditions uniquely determine the solution
describing the two closed string final states.

We shall determine the solution describing the two outgoing solutions
in two different ways: 

\smallskip

\noindent a) By explicitly finding the Fourier modes. This will show, in particular,
that the outgoing strings are in a highly excited state which is not of
maximum angular momentum.

\noindent b) By a direct matching with the original solution and imposing the new
periodicity. By construction, the resulting solutions
coincide with the explicit calculation of a), and thus this method provides a 
simple closed analytic formula for the
resummation of the Fourier expansion. This form of the solution also exhibits the
nature of the motion of the outgoing strings I and II.

\subsection{Fourier analysis}

Let us consider the string I. To find the explicit solution,
 we start with the general solution to the string equations 
for the closed string. The condition (\ref{cdoz}) can be equivalently imposed 
in the interval $-{a\pi\over 2} >\s> {a\pi\over 2}$,
since the original solution is periodic in $\s $ with period $\pi $.
The most general closed string solution satisfying the periodicity condition 
$X^\mu_\I(\s+\pi a)=X^\mu_\I(\s )$ is given by
\beq
X_\I (\s,\tau )=x_{0\I}+ 2\a' p_\I ^x\ {\tau \over a}+
i \sum_{n\neq 0} \bigg( x_n e^{-2i {n\over a}(\tau -\s )}+
\td x_n e^{-2i {n\over a}(\tau +\s )} \bigg) \ ,
\label{ffz}
\eeq
\beq
Y_\I (\s,\tau )=y_{0\I}+ 2\a' p_\I ^y\ {\tau \over a} +
i \sum_{n\neq 0} \bigg( y_n e^{-2i {n\over a}(\tau -\s )}+
\td y_n e^{-2i {n\over a}(\tau +\s )} \bigg) \ .
\label{ggz}
\eeq
Note that the constraints
$$
\dot X\cdot X' =0\ ,\ \ \ \ \dot X \cdot \dot X + {X'\cdot X'}=0\ ,
$$
are satisfied automatically once we impose the initial conditions. 
Indeed,  they are satisfied at $\tau =0$, because there
the solutions $X_I^\mu, \ X_{II}^\mu$ and their first derivatives
coincide with the original solution $X^\mu $, which already satisfies the constraints.
Since they are a constant of motion, then they are satisfied for all $\tau $
(using the fact that $T_{++}=T_{++}(\s^+)\ ,\ \ T_{--}=T_{--}(\s^-)$, one sees that
the condition $T_{++}=T_{--}=0$ at $\tau=0$ implies that they vanish for any $\tau $).

 Let us first 
 determine the conserved quantum numbers, energy, linear momentum and
angular momentum of each string. Since they are conserved 
quantities, they can be found at $\tau =0$, where the solution is 
given by eq.~(\ref{cdoz}), (\ref{cunos}). The energy  and  linear 
momentum components are given by 
\be 
E_\I &=& {2\over 2\pi \a' 
}\int_0^{{\pi a\over 2} } d\s\ \dot X^0_\I ={La\over \a' }\ ,
\\ 
p_\I^x &=& 0\ ,\ 
\\
p_\I^y &=& {2\over 2\pi \a' } 
\int_0^{{\pi a\over 2}}d\s\ \dot Y_\I={4L\over 2\pi \a'} 
\int_0^{{\pi a\over 2}} d\s\ \cos(2\s )={L\sin(\pi a)\over \pi 
\a' }\ . 
\ee
There is an extra factor of two in the above expressions, which takes into account 
that there are two segments of string giving the same contribution.
The angular momentum is 
\beq 
J_\I={2\over 2\pi 
\a'}\int_0^{{\pi a\over 2} }(X_\I\dot Y_\I-\dot X_\I 
Y_\I)={L^2a\over 2 \a'}\ \left (1+{\sin(2\pi a)\over 2\pi 
a}\right)\ . 
\eeq 
This has an orbital component $l_\I $ and spin 
component $S_\I$. Since $p^x_\I=0$, the orbital component is just 
$l_\I=x_{0\I}p^y_\I$, where $x_{0\I}$ is the center of mass 
coordinate of the string I, 
\beq 
x_{0\I}={2\over \pi 
a}\int_0^{{\pi a\over 2}} d\s\ X_\I ={L\sin(\pi a)\over \pi a 
}={\a' \over a}\ p^y_\I \ .
\eeq 
Therefore $J_\I=l_\I+S_\I$ with 
\beq 
l_\I={L^2a\over \a' }\ {\sin^2(\pi a)\over (\pi a)^2} \ ,\ \ \ \ 
S_\I ={L^2a\over 2 \a'}\ \left (1- {2\sin^2(\pi a)\over (\pi a)^2} 
+{\sin(2\pi a)\over 2\pi a}\right) \ .
\label{orbe}
\eeq 
The mass of the string I 
is thus given by 
\beq 
M^2_\I= E_\I^2- p_\I^2={L^2\over  {\a'}^2} 
\left( a^2 - {\sin^2(\pi a)\over \pi ^2} \right) \ .
\eeq 

Let us now 
determine the oscillator modes. {}From the conditions $\dot X_\I 
(\s ,0)=\dot X (\s, 0)=0$, $Y_\I (\s ,0)=Y (\s, 0)=0$, it follows 
that 
\beq 
x_n=\td x_{-n}\ ,\ \ \ \ \ y_n=-\td y_{-n}\ .\ 
\eeq 
To 
find $x_n$, $y_n$, we multiply the two remaining boundary conditions $ X_\I (\s 
,0)= X (\s, 0)$, $\dot Y_\I (\s ,0)=\dot Y (\s, 0)=0$ by 
$e^{-2i{n\over a}\s }$ and perform the integral over $\s $ from 
$-{a\pi\over 2}$ to ${a\pi\over 2}$ , using the expansions 
(\ref{ffz}), (\ref{ggz}) and the solution (\ref{ccero}). We find 
\beq 
X_\I={L\sin(\pi a)\over \pi a} \left( 1+2\sum_{n=1}^\infty 
{(-1)^n\over 1-{n^2\over a^2} } \cos({2n\tau\over a})  
\cos({2n\s\over a}) \right) \ ,
\label{solux} 
\eeq 
\beq 
Y_\I={L\sin(\pi a)\over \pi a} \left( 2\tau +2a \sum_{n=1}^\infty 
{(-1)^n\over n(1-{n^2\over a^2}) } \sin({2n\tau\over a})  
\cos({2n\s\over a}) \right) \ ,
\label{soluy} 
\eeq 
where $-{\pi a\over 2}<\s<{\pi a\over 2}$ (note that the factor 
$(-1)^n$ can be removed byshifting $\s $ so that
 $0<\s<{\pi a}$).
Finally, we have 
$X^0_\I=2L\tau $.

Let us now consider the string II.
The solution is readily found by noting that the Fourier analysis become the same in terms of
 $a'=1-a$ and $\s'=\s-{\pi\over 2}$.
We have to take into account that the shift in $\s $ produces a change of sign in the solution 
(\ref{ccero}).
We get
\beq
X_\II=-{L\sin(\pi a)\over \pi (1-a)} \left( 1+2\sum_{n=1}^\infty {(-1)^n\over 1-{n^2\over (1-a)^2} } 
\cos({2n\tau\over (1-a)})  \cos({2n\s\over (1-a)}) \right)
\label{dsolux}
\eeq
\beq
Y_\II=-{L\sin(\pi a)\over \pi (1-a)} \left( 2\tau +2(1-a) \sum_{n=1}^\infty 
{(-1)^n\over n(1-{n^2\over (1-a)^2}) } \sin({2n\tau\over (1-a)})  \cos({2n\s\over (1-a)}) \right)
\label{dsoluy}
\eeq
with $-{\pi (1-a)\over 2}<\s<{\pi (1-a)\over 2}$, and $X^0_\II=2L\tau $.

The conserved quantities for the 
string II are
\beq 
E_{\II } 
={L(1-a)\over \a' }\ ,\ \ \ \ p_{\II }^y=-{L\sin(\pi a)\over \pi 
\a' } \ ,
\eeq 
\beq 
J_{\II }=l_{\II }+S_{\II }\ ,\ \ \ \  
l_{\II }={L^2(1-a)\over \a' }\ {\sin^2(\pi a)\over (\pi (1-a))^2} \ 
,\ 
\label{orbis}
\eeq 
\beq  
S_{\II }={L^2(1-a)\over 2 \a'}\ \left (1- 
{2\sin^2(\pi a)\over (\pi (1-a))^2} - {\sin(2\pi a)\over 2\pi 
(1-a)}\right) \ ,
\eeq 
\beq M^2_{\II }= E_\II^2- p_\II^2={L^2\over  
{\a'}^2} \left( (1-a)^2 - {\sin^2(\pi a)\over \pi ^2} \right) \ .
\eeq

One easily checks that energy, linear momentum and angular momentum
are conserved in the process of splitting,
\beq
E_\I+E_\II ={L\over\a '}=E\ ,\ \ \ \ p_\I^y+p_\II^y=0\ ,\ \ \ \ \ 
J_\I+J_\II={L^2\over 2\a'}=J\ .
\eeq
We stress that the outgoing strings represent excited string states which do not 
have maximum angular momentum.


\medskip

{}For completeness, we also give the results in the case of open strings. 
The solutions are simply obtained by the formal substitution $2\s \to \s ,\ 2\tau\to \tau $ 
in eqs.~(\ref{ccero}), (\ref{solux}) -- (\ref{dsoluy}),
with the new $\s $ defined in the intervals $0<\s<\pi a$ 
and $0<\s<\pi (1-a)$, respectively. 
The expressions for $E_I,\ p_I^y,\ J_I,\ l_I,\ S_I$
are the same as above with an extra factor 1/2, and similarly for the string II.

\subsection{Closed formulas for the outgoing string solutions}

\def\t{\tau }

We can describe the classical closed-string dynamics by means of left and right motion,
in terms of the coordinates $\s^{\pm}=\s\pm \t $:
\beq
X^{\mu}(\s,\t )=X^{\mu}_{+}(\s^{+})+X^{\mu}_{-}(\s^{-}) \ .
\label{motion} 
\eeq
The constraint is $\eta_{\mu\nu}\de_{\pm}X^{\mu}_{\pm}\de_{\pm }X^{\nu}_{\pm}=0$.

The initial string (\ref{ccero}) 
is described by:
\beq
X_{\pm}(\s^\pm )={L\ov 2}\cos(2\s^\pm )\ ,\ \ \ Y_{\pm}(\s^\pm )=\pm {L\ov 2}\sin(2\s^\pm)\ , \ \ \ 
X^0_{\pm}(\s^\pm )=\pm {L}\s^\pm\ . 
\label{rotating}
\eeq
As before, $\s \in [0,\pi )$.  

At  $\t=0$ the string splits into two pieces:
$X^\mu _{\pm}( \s^\pm)\to \{ X^\mu_{\I\pm} ( \s^\pm ), X^\mu_{\II\pm } (\s^\pm ) \} $.
The two pieces are folded like the initial string and the splitting occurs 
at $\s={a\pi\over 2}$ in the upper segment and at $\s=\pi-{a\pi\over 2}$ in the lower segment,
with $0<a<1$. The two resulting strings are determined as in the previous subsection
by requiring continuity
of the string coordinates and their first derivatives in $\t$ at $\t=0$, and by requiring 
periodicity in $\s$:   $X^{\mu}_{\I}$ with period $\Delta \s=a\pi$ and  
$X^{\mu}_{\II}$ with period $\Delta \s=(1-a)\pi$.

In the interval of $\s $ corresponding to string I
(see eq.~(\ref{cdoz})~), the initial conditions at $\tau =0$ imply that 
$\partial_\s X(\s,0 )=X_{+}'(\s)+X_{-}'(\s )=X_{\I+}'(\s )+X_{\I-}'(\s )$ and 
$\partial_\tau X(\s,0 )=X_{+}'(\s )+X_{-}'(\s )=X_{+}'(\s )-X_{-}'(\s )$, which in turn imply that
$X_{I+}(\s )=X_+(\s)$ and $X_{\I-}(\s )=X_{ -}(\s)$ 
in this interval, and outside of the interval they are defined by the 
new periodic boundary condition $\s\to\s+ a\pi $.
This determines $X_{I}(\s,\tau )= X_{I+}(\s^+)+X_{I-}(\s^-)$ in closed form.
 Similarly for the coordinate $Y$, and for the string II.

The resulting expressions for $Y_{\I,\II}$ are the sum of two terms.
One term corresponds to the momentum carried by the string:  
$\pm L{2\sin(a\pi )\ov\pi}\tau$. 
The other term is periodic in $\s$ and its derivative in $\s$ has zero average.

It is convenient to rescale the world-sheet parameters of the resulting strings
$\s^ {\pm}\to a \s^{\pm}$ for the string $\I$, 
and $\s^ {\pm}\to (1-a) \s^{\pm}$ for
the string $\II$, in such a way that the period is $\Delta \s =\pi$
for both. 

We get for the string $\I$:
\be
X^0_{\I\pm}(\s^\pm)&=&\pm {L}a\s^\pm \ ,
\nonumber \\
X_{\I\pm}(\s^\pm)&=&{L\ov 2}C_\I(\s^\pm ) \ , \
Y_{\I\pm}(\s^\pm )=\pm {L\ov 2}\big[ {2\sin(a\pi )\ov\pi}\s^\pm +S_\I(\s^\pm )\big]
\label{stringauno}
\ee
where
\be
&C_\I(\s )&=\cos(2a\s ) \ , \ \ S_\I (\s )=\sin(2a\s )-{2\sin(a\pi )\ov\pi}\s  \nonumber\\
&{\rm for}& \ \  0\leq \s < {\pi\over 2} \ , \nonumber \\
&C_\I(\s )&=\cos(2a\s-a2\pi) \ , \ \ S_\I(\s )=\sin(2a\s-a2\pi )-{2\sin(a\pi )\ov\pi}(\s -\pi )
 \nonumber\\
&{\rm for}&  \ {\pi\over 2}\leq \s < \pi\ . 
\ee

Similarly for the string $\II$ we get:
\be
&&X^0_{\II\pm}(\s^\pm )=\pm {L}(1-a)\s^\pm \ ,
\nonumber \\
&&X_{\II\pm}(\s^\pm )={L\ov 2}C_{\II}(\s^\pm )  , \
\nonumber\\
&&Y_{\II\pm}(\s^\pm )=\pm {L\ov 2}\big[ -{2\sin(a\pi )\ov\pi}\s^\pm+S_{\II}(\s^\pm )\big]  
\label{stringadue} 
\ee
where
\be
&&C_{\II}(\s )=\cos(2(1-a)\s + a\pi) \ , \ \
\nonumber\\ 
&&S_{\II}(\s )=\sin(2(1-a)\s + a\pi)+{2\sin(a\pi )\ov\pi}\s 
\nonumber\\
 &&{\rm for} \  \ 0\leq \s < \pi \ .
\ee
These definitions are extended to any $\s$ by declaring that  
$$
C_{\I,\II}(\s +\pi )=C_{\I,\II }(\s )\ ,\ \ \ \ 
S_{\I ,\II } (\s +\pi )=S_{\I,\II}(\s )\ .
$$
They solutions are equivalent to the solutions given by 
the Fourier expansion in the previous section.

The derivative in $\s $ of both $X_{\I},Y_{\I}$ and $X_{\II},Y_{\II}$ 
has a discontinuity
at $\s^\pm =\pi/2$ and $\s^\pm =0$ respectively. 
This discontinuity will appear as an angular bending  in the
(folded) shape of the strings $\I$ and $\II$. Since $\s^{\pm}=\s\pm \t$, this angular
bending will
move along the string, as a function of $\tau $.
We will return to this point in the last section.

\section{Comparison with the superstring  quantum calculation}

The decay process described above by semiclassical splitting predicts that masses will 
be related by the following formulas:
\beq 
M_\I= M_\I (a)= {L\over  {\a'}} \sqrt{a^2 - {\sin^2(\pi 
a)\over \pi ^2} }\ ,
\label{asd}
\eeq 
\beq M_{\II }=M_{\II}(a ) ={L\over  
{\a'}} \sqrt{ (1-a)^2 - {\sin^2(\pi a)\over \pi ^2} }\ .
\label{zxc}
\eeq
These relations define a function $M_\I=M_\I(M_\II )$.
 
In ref.~\cite{IR},
the full quantum calculation of the decay was done in the ten dimensional type II superstring theory.
The decay rate is obtained by extracting the imaginary part of
the genus one self-energy of the massive particle:
$\Gamma ={\rm Im}{\Delta M^2\over 2M}$. 

This calculation is complicated: it combines expansions of theta functions and resummations, 
saddle-point   evaluation of some integrals. Nevertheless, it is an exact genus one result 
in the large $M$ limit, since the only approximation involved are saddle-point approximations, 
which become exact as $\a'M^2\gg 1$. 

Here we explain briefly the idea of the method and the result. 
We compute the contribution to ${\rm Im}(\Delta M^2)$ of the decay channel 
corresponding to the states with masses $M_1, M_2$: let us call  
${d^2{\rm Im}(\Delta M^2)\ov dM^2_1dM^2_2}$ that contribution per unit  $ dM^2_1dM^2_2$.
The one-loop self-energy, derived first in \cite{IK}, 
is represented as an integral of some combination of theta functions, 
the integration being over the torus complex modulus and over a complex 
(vertex position) coordinate on the torus surface. The integration is formally divergent,
and it is computed by a standard analytic continuation. 

A key point is to write the integrand as a sum of holomorphically factorized quantities.
The selection of a particular decay channel corresponds to a particular
 term in the Taylor expansion of the holomorphic factors, as it is recognized
by comparison with field theory Feynman diagrams. 
Cauchy contour integrals and  saddle point techniques 
are used for getting the holomorphic expansion, similarly to the well known
procedure for computing the entropy of a state in string theory.    
In this way we obtained:
\beq
{d^2{\rm Im}(\Delta M^2)\ov dM^2_1dM^2_2}\sim g_s^2\ 
M^{-3}\exp[2M^2 S_0(M_1/M,M_2/M)]\ ,
\label{quantum}
\eeq
with $S_0\leq 0$ (given in fig.~1 of \cite{IR}).

Therefore in the large mass limit the dominant decay channel corresponds to values 
of $M_{1,2}$
for which $S_0=0$, i.e. the masses of the 
decay products are correlated, modulo processes which are 
exponentially suppressed. 
This effect may come as a surprise, 
since one might have expected a sizable string vertex coupling
three string states of arbitrary masses $M,\ M_1,\ M_2$.

Figure 1 shows the relation $M_\I=M_\I^{\rm quantum }(M_\II )$, 
corresponding to $S_0=0$, found numerically in \cite{IR} (and shown there in fig.~3).
. 
In the same figure, 
we have superposed the analytic function $M_\I=M_\I(M_\II )$ 
defined by (\ref{asd}), (\ref{zxc}).
We see that they 
coincide, the analytic semiclassical curve fully matches the curve of \cite{IR}
obtained by a one-loop calculation.

\bigskip

\begin{figure}[hbt]
\label{fig1} 
\vskip -0.5cm \hskip -1cm
\centerline{\epsfig{figure=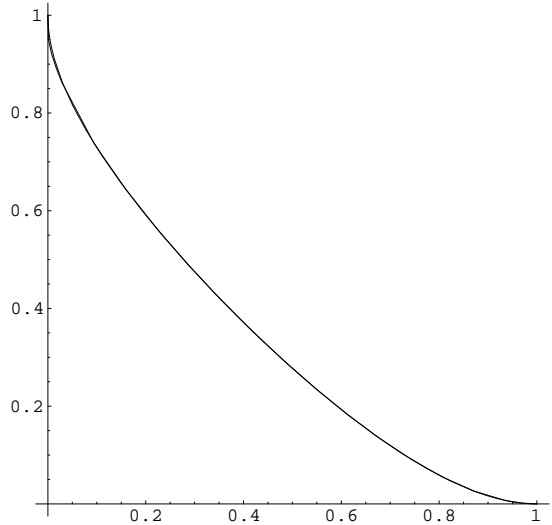,height=7truecm}}
%
\caption{\footnotesize 
The curve $M_\I =M_\I (M_\II)$ defined by eqs.~(\ref{asd}), (\ref{zxc}).
It is superposed with the similar curve obtained in \cite{IR} from the direct
quantum one-loop calculation.}
\end{figure}

This precise match is also surprising.
Although one expects that for a  large mass  the initial highly 
excited quantum string state is well described by a classical solution, 
it is not a priori obvious that the precise relation 
between masses should be implied  by a classical spontaneous splitting process.
Moreover, the decay products are not states of maximum angular momentum and it is not obvious 
that a semiclassical description would be applicable for them. 
The accurate coincidence of the two curves in figure 1 also confirms the results of \cite{IR}.


Further, in the calculation of \cite{IR}, it is also possible to isolate the contribution 
of a given orbital angular momentum $l_0$. In fact, by
comparison with the field theory expressions for one-loop 
Feynman diagrams, one learns that the sum over  the possible
 $l_0$ contributions  corresponds to the sum over the holomorphically factorized terms
(in eq. (5.4) of \cite{IR}, a term with given $l_0$ has 
$l_0=2N-(m_1+m_2)-2={\rm fixed},\ N=J/2$).

However, $l_0$ is not a quantum number of the final states,
and in fact the quantum computation of ${\rm Im}(\Delta M^2)$ is expressed as the modulus 
square of a sum over amplitudes which are alternating in sign.
Therefore, the classical value of $l_0$ does not correspond to a well defined
mean value.  In order to compare the quantum computation with the classical
configuration of outgoing states $M_1,M_2$ with given $l_0$, 
we have taken as quantum value of $l_0$  the one for which
the amplitude is maximal in absolute value.
That maximal amplitude corresponds  to a value of $l_0$ which depends on $M_1,M_2$. 
We take the relevant values for $M_{1,2}$ to be those along the curve of
Figure 1,  and then compare this result to $l_\I(a)+l_\II (a)$ 
computed semiclassically (see eqs. (\ref{orbe}), (\ref{orbis})~). 
This comparison is shown in Figure 2. 

\bigskip

\begin{figure}[hbt]
\label{fig2} 
\vskip -0.5cm \hskip -1cm
\centerline{\epsfig{figure=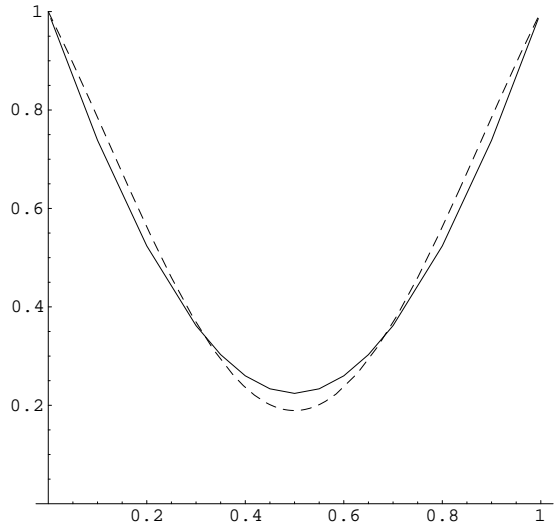,height=7truecm}}

\caption{\footnotesize
Plot of the classical (dashed line) and quantum (solid line) value for
$(1-l_0/J)$ ($l_0$ = orbital angular momentum) 
as a function of the parameter 
$a$ (which defines the breaking point of the string). }
\end{figure}


We see that the two curves are  quite close.
We think that the small discrepancy is due 
to the fact that there is not a well defined quantum value of $l_0$. 
It is still remarkable that the semiclassical and quantum calculation give  
so close results, with the correct normalization which comes 
automatically without any tuning.

\section{Decay rate and description of the motion}

{}For quantum states with large occupation numbers  admitting a semiclassical
description, one expects for the lifetime an expression of the form
\beq
{\cal T}=  \Gamma^{-1}= {\rm const.} {1\over g_s^2}\ {\cal T}_0 \ \exp[-2{L^2\ov \a'}S_0^{Max}]\ .
\label{tric}
\eeq
This is precisely the form of the result (\ref{quantum}) 
that arises from the explicit quantum calculation.
We have seen that in general $S_0\leq 0$ and that the maximum of $S_0$  
corresponds to the classical solution for which $S_0^{Max}=0$.
The total contribution to $\Gamma $ is obtained by performing
the integral over  $dM_1^2,\ dM_2^2$ of (\ref{quantum}) \cite{IR}. 
Only a small neighborhood
around the curve of fig.~1 contributes, since other regions are suppressed
exponentially. 
Using the semiclassical form (\ref{quantum}) of $S_0$ and expanding
in the vicinity of the curve, 
one is left with a Gaussian integral in the orthogonal direction
of the curve which produces an additional factor of $1/\sqrt{N}$. 
This procedure gives (\ref{tric}) with ${\cal T}_0 \cong L$, 
i.e. ${\cal T}= \Gamma^{-1}= {\rm const.}\ {\a '\over g_s^2}\ M$,
which was the result reported in \cite{IR}.
However, while the saddle point approximation gives an accurate formula
for the 
exponential part of the decay rate, obtaining the 
power behavior is subtle in the present case, 
due to the vanishing of some determinants in the vicinity
of the curve of fig.~1. The correct power behavior is being considered
in a work in progress.

\medskip

Having the exact solutions for the outgoing strings I and II, 
it is interesting to describe the main features of their motion.
Figures 3 and 4 are  plots of a sequence of pictures of the string I 
and II after the splitting, for different breaking points: $a=0.4$, $a=0.15$ respectively.
We see that the outgoing closed string remain folded, exhibiting a rotating motion.
The figures 3 and 4 were made using the Fourier series formulas of section 2.1.
One can check  that the same plot is obtained using the
formulas of section 2.2.
Figures 5,6,7 are a plot of the world-sheet for the breaking point at $a=0.4$.

The most salient feature that can be observed from the figure is that  
the breaking of the strings creates  an angular bending, or kink, which 
then travels back and forth all along each string.
Remarkably, each string is straight except at the bending point.
One might wonder whether this feature is generic, at least 
for the breaking of an open string:
the kink is produced by the jump of the first derivative at the splitting point; it
is locally created, thus its
occurrence should not depend on whether the string has maximum angular momentum 
(i.e. whether it is straight or  curved).

Another important feature is that the angles of the bendings of the strings I and II
sum up $180^{\rm o}$, and they are given by
\beq
\theta_\I = a\pi\ ,\ \ \  \theta_\II=(1-a) \pi \ .
\label{angulos}
\eeq 
Note that they are in relation with the energies of the strings.
When $a=1/2$  both angles are $\pi/2$.

The formulas (\ref{angulos}) can be proved by using the solutions of section 2.2.
Consider the string I at a given instant $\tau $. 
The derivative ${dY_\I\over dX_\I}$ has the same discontinuity in both upper 
and lower segments of the closed string. In one segment the discontinuity
 originates from the discontinuity in $\partial_\s X_{\I +}$ at 
$\s_1 ={\pi\over 2}-\tau $. In the other segment, it originates 
from the discontinuity of 
$\partial_\s X_{\I -}$ at $\s_2 ={\pi\over 2}+\tau $.

{}For $a<1/2$, the bending angle $\theta_I$ of the string I is an acute angle.
It can be computed in particular at $\s_1 $, where it is given by
\beq
\theta_\I=\arctan{dY_\I\over dX_\I}\bigg|_{\s_1-\epsilon}
 -\arctan{dY_\I\over dX_\I}\bigg|_{\s_1+\epsilon} \ .
\eeq
Using the explicit form of the solutions, we obtain
$$
{dY_\I\over dX_\I}\bigg|_{\s_1-\epsilon} = 
{\partial_\s Y\over \partial_\s X}\bigg|_{\s_1-\epsilon}=\tan(2a \tau)\ ,
$$
\beq
{dY_\I\over dX_\I}\bigg|_{\s_1+\epsilon} = \tan(2a \tau-a\pi  )\ .
\eeq
Thus
\beq
\theta_\I=2a\tau-(2a\tau-a\pi)  =a\pi\ .
\eeq
Similarly, one finds $\theta_\II $ as in (\ref{angulos}).

\bigskip
\bigskip

\noindent {\bf Acknowledgements:} 

We thank J. Maldacena for suggesting the basic idea that led to this investigation, 
namely
that the relation between the masses $M_1,M_2$ of the decay products found in 
\cite{IR} could be explained in terms of spontaneous splitting 
of the rotating string solution. R.I. acknowledges partial support 
by the EEC grant HPRN-CT-2000-00131.


\begin{figure}[hbt]
\label{fig3} 
\vskip -0.5cm \hskip -1cm
\centerline{\epsfig{figure=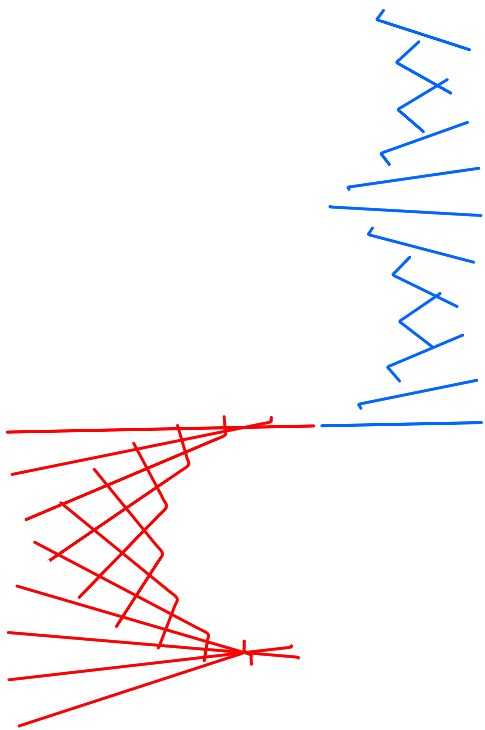,height=10truecm}}
\caption{\footnotesize
Sequences of pictures of the closed string after the splitting for $a=0.4$. }
\end{figure}

\begin{figure}[hbt]
\label{fig4} 
\vskip -0.5cm \hskip -1cm
\centerline{\epsfig{figure=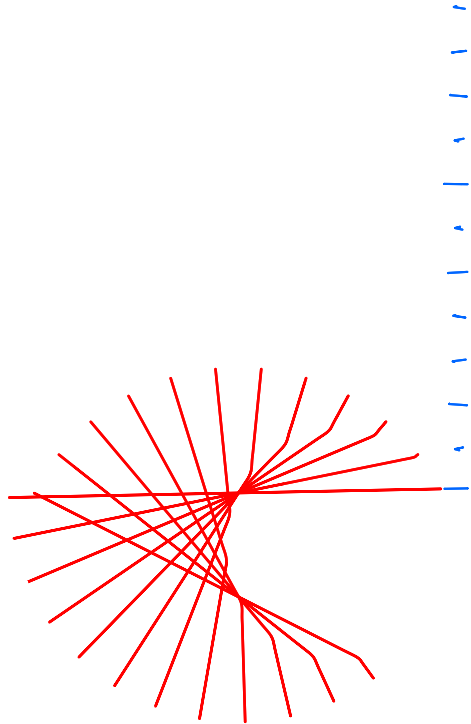,height=10truecm}}
\caption{\footnotesize
Sequences of pictures of the closed string after the splitting for $a=0.15$. 
One can see the slow translational motion of the large string, which is a
small deformation of the original rotating string solution, with a bending angle
$0.85\pi $. The small string moves very fast and the bending angle is acute,
equal to $0.15 \pi $.}
\end{figure}

\begin{figure}[hbt]
\label{fig6} 
\vskip -0.5cm \hskip -1cm
\centerline{\epsfig{figure=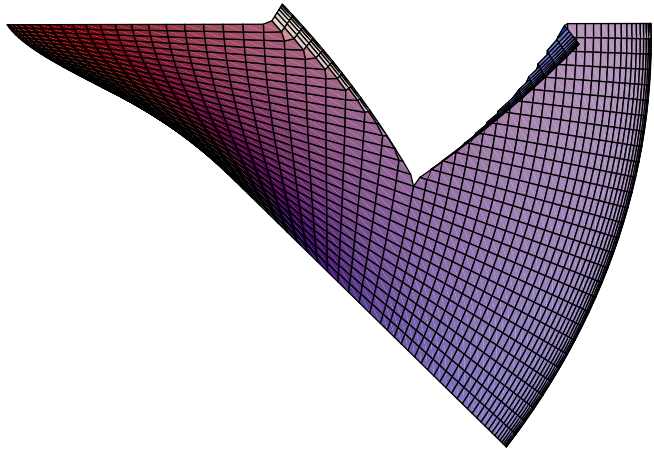,height=10truecm}}
\caption{\footnotesize
Surface swept by the strings during the splitting process (lateral view) . }
\end{figure}

\begin{figure}[hbt]
\label{fig7} 
\vskip -0.5cm \hskip -1cm
\centerline{\epsfig{figure=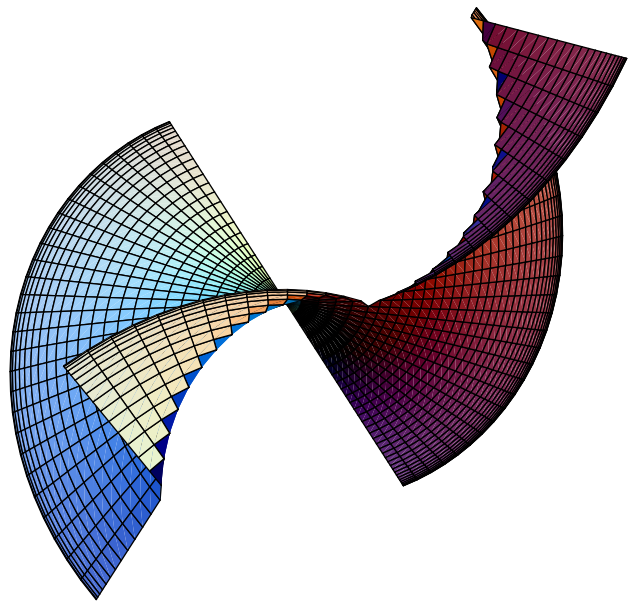,height=10truecm}}
\caption{\footnotesize
Surface swept by the strings during the splitting process (view from the top, larger time interval) . }
\end{figure}

\begin{figure}[hbt]
\label{fig8} 
\vskip -0.5cm \hskip -1cm
\centerline{\epsfig{figure=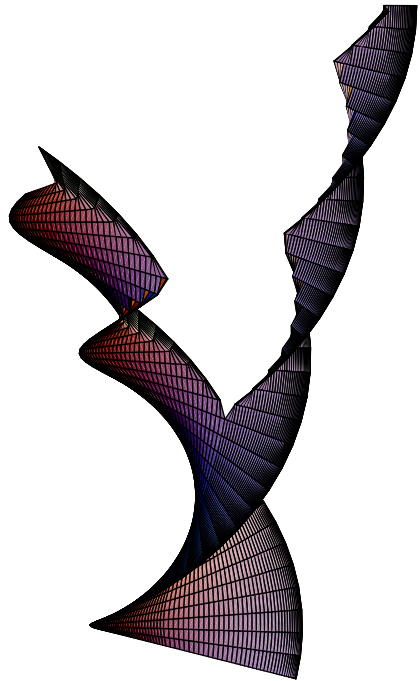,height=12truecm}}
\caption{\footnotesize
Surface swept by the strings during the splitting process (lateral view, even larger time interval) . }
\end{figure}


 \end{document}